\documentclass[aps,prd,10pt,twocolumn,preprintnumbers,nofootinbib,superscriptaddress,a4paper,showpacs]{revtex4-1}
\usepackage{graphicx}
\usepackage{amssymb}
\usepackage{color}
\usepackage{enumerate}
\usepackage[OT2,OT1]{fontenc}
\usepackage{natbib}
\newcommand\cyr{%
\renewcommand\rmdefault{wncyr}%
\renewcommand\sfdefault{wncyss}%
\renewcommand\encodingdefault{OT2}%
\normalfont
\selectfont}
\DeclareTextFontCommand{\textcyr}{\cyr}

\def\beq{\begin{equation}}
\def\eeq{\end{equation}}

\def\be{\begin{eqnarray}}
\def\ee{\end{eqnarray}}

\renewcommand{\texttt}{{}}

\renewcommand{\texttt}{{}}

\def\bs{\begin{subequations}}
\def\es{\end{subequations}}
\def\a{\alpha}

\def\de{\delta}

\def\cL{\mathcal{L}}

\def\cS{\mathcal{S}}

\def\ds{d_{\rm S}}

\def\p{\partial}

\def\B{\Box}
\newcommand{\Eq}[1]{(\ref{#1})}
\def\com{\color{magenta}}
\def\cob{\color{blue}}
\newcommand{\book}[5]{\emph{#1} (#2, #3, #4, #5)}
\newcommand{\oarX}[1]{\href{http://arxiv.org/abs/#1}{{\ttfamily\com arXiv:#1}}}
\newcommand{\arX}[1]{\href{http://arxiv.org/abs/#1}{{\ttfamily\com arXiv:#1}}}
\newcommand{\doin}[6]{\href{http://dx.doi.org/#1}{{\cob #2 #3 {\bf #4}, #5 (#6)}}}
\newcommand{\doinn}[5]{\href{http://dx.doi.org/#1}{{\cob #2 {\bf #3}, #4 (#5)}}}
\newcommand{\doij}[5]{\href{http://dx.doi.org/#1}{{\cob #2 {#3} (#5) #4}}}

\newcommand{\ndoinn}[5]{\href{#1}{{\cob #2 {\bf #3}, #4 (#5)}}}
\newcommand{\tia}[1]{}

\def\rme{e}
\def\rmd{d}
\def\rmi{i}

\begin{document}

\title{Nonlocal quantum gravity and M-theory}
%
\author{Gianluca Calcagni}
\thanks{Electronic address: calcagni@iem.cfmac.csic.es}
\affiliation{Instituto de Estructura de la Materia, CSIC, Serrano 121, 28006 Madrid, Spain}
\author{Leonardo Modesto}
\thanks{Electronic address: lmodesto@fudan.edu.cn}
\affiliation{Department of Physics \& Center for Field Theory and Particle Physics, 
 Fudan University, 200433 Shanghai, China}

\date{\small April 7, 2014}

\begin{abstract} \noindent
We construct a ultraviolet completion of the bosonic sector of 11-dimensional supergravity motivated by string field theory. We start from a general class of theories characterized by an entire nonpolynomial form factor that allows one to avoid new poles in the propagator and improves the high-energy behavior of the loops amplitudes. Comparing these models with effective string field theory, a unique form factor is selected out. In view of this, we modify ten-dimensional supergravity and finally get a ultraviolet completion of 11-dimensional supergravity by an oxidation process. The result is a candidate for a finite and unitary particle-field limit of M-theory.

\end{abstract}
\pacs{04.60.-m, 11.10.Lm, 11.25.Yb}

\preprint{\doin{10.1103/PhysRevD.91.124059}{PHYSICAL REVIEW}{D}{91}{124059}{2015} \hspace{9cm} \arX{1404.2137}}


\maketitle

\section{Introduction}\label{intro}

The particle-field limit of string field theory (SFT) and the properties of its gravity sector are territories still subject to exploration. Having assumed that one has integrated out all massive degrees of freedom of string theory, what is the form of the resulting massless quantum field theory and what are its ultraviolet (UV) properties? To give a definite answer to such a question, one should calculate an effective Wilsonian action for the massless fields. The form of the action should then be fixed by deriving an exact renormalization-group equation that determines the explicit cutoff dependence of the infinite set of terms. In particular, one possibility would be to take the Batalin--Vilkovisky equation rigorously obtained in string field theory \cite{Zwi92,HaZw} and reduce its solution to an exact Wilsonian action for the massless fields. This challenge is rather demanding from a technical point of view and, to the best of our knowledge, still unaccepted. However, to see the type of result one might expect to obtain, we follow a different path. 

In this paper, we explore the possibility to achieve a super-renormalizable theory of gravity in ten dimensions capable of capturing the main properties of effective string field theory. In this respect, and to summarize the steps of the strategy employed here, we will start from a class of nonlocal quantum-gravity models that are independent of string theory. These models have generic but well-behaved (in a sense made precise below) form factors, i.e., nonlocal covariant operators with an infinite number of derivatives. The next step is to fix such form factors uniquely by requiring the linearized action for the graviton to match that of string field theory. The result is a ten-dimensional field theory for gravity which is proposed to describe simultaneously the nonlinear (in the metric) extension of the closed-SFT effective nonlocal action for the graviton and the nonlocal extension of ten-dimensional supergravity. The third step consists of asking ourselves whether we can further extend the model to 11 dimensions, what its properties are, and whether these properties are compatible with the known symmetries and spectrum of M-theory. From the $D=10$ nonlocal model, by  an ``oxidation'' process from 10 to 11 dimensions, we are able to get a finite completion of the bosonic sector of 11-dimensional supergravity. After some checks of the properties of the resulting $D=11$ action, we propose it as a nonlocal-field-theory limit of M-theory.\footnote{In general, there is some margin of ambiguity about what M-theory means. According to modern notions, it is an 11-dimensional theory with a membrane population that also possibly admits a formulation as a matrix model at the quantum level. By ``M-theory,'' one also refers to the web of dualities among different superstring theories and between these theories and 11-dimensional supergravity. The latter is regarded as the limit where the light degrees of freedom of M-theory can be described by a (quantum) field theory. A nonlocal version of $D=11$ supergravity is the main focus of the present paper.}

To this purpose, we combine results found since the 1980s in string theory \cite{Wi86a,W11,W2}, earlier and parallel studies on nonlocal quantum field theory by Efimov \cite{AE1,AE2,Efi77,Efi01}, Krasnikov and Tomboulis \cite{Kra87,Tom97}, and Moffat and Cornish \cite{Mof1,HaM,EMKW,Cor1,Cor2,Cor3,Mof3}, and several more recent ideas in quantum gravity \cite{BMS, cuta8, BKM1,Mod1, BGKM, Mod2, Mod3, Mod4, Mod5, BKMV} (see also Refs.\ \cite{odi, Deser1, Deser2, Modesto:2013jea, Maggiore} for nonlocal infrared modifications to gravity). The theory here proposed aims to fulfill a synthesis of minimal requirements:
\begin{itemize}
\item[(i)] Spacetime is a continuum in which Lorentz invariance is preserved at all scales.
\item[(ii)] Classical local supergravity should be a good approximation at low energy.
\item[(iii)] The theory has to be perturbatively super-renormalizable or finite at the quantum level (consequently, the effective spacetime dimension decreases with the energy).
\item[(iv)] The theory has to be unitary, without extra degrees of freedom in addition to those present in the classical theory.
\item[(v)] Typical classical solutions must be singularity free.
\item[(vi)] The proposal should be manifestly compatible with the known features of string theory.
\end{itemize}

Actually, there is an infinite class of theories compatible with properties (i)--(v), parametrized by different nonpolynomial entire functions. We will see that, in order to agree with string field theory and fulfill (vi), the entire function will be singled out {\em uniquely}. The $D=11$ bosonic sector of the Lagrangian we are going to construct step by step reads as
\be
 \mathcal{L}_{11} &=& \frac{1}{2 \kappa_{11}^2 } \!
\left(R + G_{\mu \nu } \frac{\rme^{ - \Box_{\Lambda} } -1}{\Box} R^{\mu \nu } - \frac{1}{2 \times 4 !} F_4  \rme^{ - \Box_{\Lambda} }  F_4 \right) \nonumber \\
&& + \cL(A_3) +\cL(\psi)\,, 
\label{1}
\ee
where $F_4$ is the field strength of the 3-form $A_3$ and $\cL(A_3)$ and $\cL(\psi)$ are nonpolynomial modifications of the Chern--Simons and fermionic actions. The metric tensor $g_{\mu \nu}$ has signature $(- ,+, +, \dots, +)$; the curvature tensor, Ricci tensor and curvature scalar are defined by, respectively, $R^{\alpha}_{\ \beta \gamma \delta} = - \partial_{\delta} \Gamma^{\alpha}_{\beta \gamma} + \dots$, $R_{\mu \nu} = R^{\alpha}_{\ \mu \alpha \nu}$, and $R = g^{\mu \nu} R_{\mu \nu}$. $G_{\mu\nu}=R_{\mu\nu}-(1/2) g_{\mu\nu} R$ is the Einstein tensor. The operator $\Box_{\Lambda}=\B/\Lambda^2$ is the d'Alembertian divided by an energy scale
 $\Lambda$.
 In momentum space, $- \Box_{\Lambda} \to k^2 \to  k_{\rm E}^2$, where in the second step we transformed to Euclidean momentum (hence the index ``E''), $k_0 \rightarrow - \rmi k_4$.

The form of the kinetic terms in the particle-field-theory limit of SFT fixes the form factor of the above gravity models uniquely among the infinitely many belonging to this class.  This, of course, is a much lesser achievement than uniquely fixing the UV complete particle-field limit of M-theory, but it indicates in a clear way the form the gravitational sector should take if ten-dimensional SFT is valid and, at the same time, if UV finiteness is demanded. 

Our \emph{Ansatz} \Eq{1} has a number of properties meeting with requirement (vi), all of which constitute important checks of its compatibility with string theory:
\begin{itemize}
\item[(a)] Contrary to local theories, the UV behavior does not get worse with dimensional reduction: both unitarity and renormalization survive after compactification (Sec.\ \ref{5to4d}).
\item[(b)] Type-II string amplitudes are recovered in the local limit (Sec.\ \ref{coch}).
\item[(c)] There exists an S-duality linking 11-dimensional (11D) M-theory to ten-dimensional (10D) Type-IIA nonlocal supergravity (Sec.\ \ref{sdua}).
\item[(d)] It is also argued (although not proved) that inclusion of massive states does not spoil the UV properties of the theory (Sec.\ \ref{massi}). 
\end{itemize}

Although the model we present is a simplification of the most complete nonlocal 11D supergravity action one can write compatibly with the M-theory content, it already displays a richness of implications that local supergravity does not possess, not only from the point of view of particle physics and the mathematical structure of the theory but also of cosmology. In particular, there is the possibility (classically) that the big-bang singularity is removed and replaced by a finite bounce, near which a natural period of superinflation occurs. This has been shown in previous studies of the type of nonlocality in Eq.~\Eq{1} \cite{BMS,cuta8,BKM1,Mod1, BGKM,Mod2,Mod3,Mod4,Mod5,BKMV,SFC} (see also Refs.\ \cite{ADDG,Bar1,Bar2,HaW,kho06} for earlier works on nonlocal gravity). Moreover, since the linearized limit of the nonlinear gravitational theory does not possess ghosts or other types of obvious instabilities, it is expected to provide a reliable model of cosmic evolution all the way back to the highest densities. This is in stark contrast with effective local gravitational actions made of higher-order curvature invariants, for which their effective description breaks down at the scale of the highest-order term. Thus, the interest in a nonlocal field-theory limit for M-theory is not merely academic as the proposal would have sizable, testable consequences for the history of the early Universe.

The gauge symmetry group of string field theory is extremely large and partly unknown. It includes spacetime diffeomorphisms and supersymmetry transformations \cite{QS}, but the way in which such symmetries arise in the low-energy field-theory limit is not straightforward. For instance, to identify general coordinate transformations in closed string theory, one has to make field redefinitions order by order in the string coupling \cite{GhSe}; the same holds for Abelian transformations in the open case \cite{Dav00}. Also, fixing the gauge freedom of the full theory is necessary to compute amplitudes and S-matrix elements that, again, would require the Batalin--Vilkovisky formalism. In this paper, we do not consider the problem of constructing a well-defined string field theory from a conformal field theory representation and, from that, derive the effective action for gravity; therefore, we cannot enter in the details of the above general and important issues. Still, we offer a concrete particle-field proposal that, regardless of its ultimate validity, can help at least in stimulating more interest in the subject. To obtain a unique theory, it is sufficient to combine together three different ingredients: general covariance, the exponential falloff of the propagator, and unitarity (ghost freedom). We then ask what a completion of 11-dimensional supergravity, compatible with quantum mechanics and which cannot be excluded in a multivacuum M-theory, should look like. Regarding the assumption of general covariance, we commented above about the difficulties in understanding how diffeomorphism invariance arises in the low-energy spacetime approximation of the theory. In a worst-case scenario, it might even happen that it is not possible to write down a nonlinear \emph{and} nonlocal fully covariant action for gravity in closed SFT. However, there is no negative proof in that direction yet and, if the theory really reproduced gravity, it must satisfy general covariance in some limit. Since we are able to formulate a quantum field theory of gravity when general covariance is respected, we make such an assumption. This field theory happens to display a class of nonlocal operators that includes those of SFT, which leads us to the second point. The exponential operator is a general feature of the particle-field limit of certain string field theories, but not of others (for instance, Berkovits's nonpolynomial theory) in which the low-energy spacetime limit has never been derived while retaining nonlocality. However, several dualities between different open SFTs are known, including a mapping between supersymmetric and bosonic classical solutions \cite{Er07a,FK3}, cubic and Berkovits's supersymmetric SFTs \cite{FK08,AGM2}, and different polynomial supersymmetric SFTs \cite{Kro91,Kro92}. These results have been found only for the open string, but they are encouraging enough to put us in the position of submitting a reasonable exploratory proposal. In fact, it is known that in closed SFTs such as the one reported in the next section the same type of nonlocality arises. This is perhaps not surprising if one recalls that the exponential derivative operator is a direct consequence of the existence of surface states in the Fock space of the open-string case \cite{cuta7} and that states similar to these exist also in the closed-string case \cite{KMWa}. Therefore, our understanding of the gauge group of SFT is already sufficient to make an educated guess about the effective gravitational action. 

We minimally review string field theory and renormalizable perturbative quantum-gravity models in Sec.\ \ref{rev}. In Sec.\ \ref{5to4d}, we show in a five-dimensional example that Kaluza--Klein compactification preserves the structure of the form factors and, hence, the renormalizability and unitarity properties of the theory. The main results in $D=11$ and $D=10$ dimensions are in Sec.\ \ref{mth}, where we provide various checks of the theory as detailed above. Inclusion of massive states is discussed in section \ref{massi}. In Sec.\ \ref{fiit}, we collect some remarks on the issue of ultraviolet finiteness in string theory and quantum gravity. Section \ref{concl} is devoted to a discussion.


\section{From string field theory to quantum gravity}\label{rev}

The purpose of this section is to sketch the type of nonlocal operators arising in string field theory.

\subsection{String field theory}\label{sft}

A string field $\Psi$ is a mathematical object interacting via certain composition rules and given by the superposition of string oscillators applied to the Fock vacuum (or vacua, in the case of the closed string). The coefficients of the level expansion are particle-field modes of progressively higher mass. The interaction and precise field content depend on the type of string. Open SFT is the simplest example \cite{ohm01,FK}. The open SFT bosonic action is \cite{Wi86a}
\be\label{SFT}
S=-\frac{1}{g_o^2}\int \left(\frac{1}{2\alpha'} \Psi* Q\Psi+\frac13\Psi*\Psi*\Psi\right),
\ee
where $g_o$ is the open string coupling, $\a'$ is the Regge slope, $*$ is a noncommutative product, and $Q$ is the Becchi--Rouet--Stora--Tyutin (BRST) operator. The resulting equation of motion is $Q\Psi+\a'\Psi*\Psi=0$. The action is invariant under the infinitesimal gauge transformation $\delta\Psi=Q\Psi+\Psi*\Gamma-\Gamma*\Psi$, where $\Gamma$ is a zero-ghost-number state. The gauge group of string field theory is very large and not completely known. It includes at least worldsheet reparametrizations, spacetime diffeomorphisms, and (in the supersymmetric extensions) supersymmetry transformations \cite{QS,ORZ}.

Fixing the gauge and keeping only the tachyonic mode $\phi$, Eq.\ \Eq{SFT} gives rise to an effective spacetime action for $\phi$ that exhibits a nonlocal interaction \cite{KS1,KS2},
\be
\cS =\frac{1}{g_o^2}\int \rmd^D x \left[\frac{1}{2}\phi\left(\B+\frac{1}{\a'}\right)\phi-\frac{\rme^{3/(\a'\Lambda^2)}}{3}\tilde\phi^3\right],\label{bosa}
\ee
where
\be
\tilde\phi := \rme^{\B_\Lambda}\phi\,,\qquad \frac{1}{\Lambda^2}=\a'\ln \left(\frac{3^{3/2}}{4}\right)\approx 0.2616\,\a'\,.
\label{rst}
\ee
The value of $\Lambda$ is dictated by conformal invariance (which is partly broken by level truncation). The other fields of the level expansion, as well as the $p$-adic model \cite{W2}, display the same type of nonlocality.

While there is only one proposal for the bosonic open string, there are many for open superstring field theory. The first by Witten \cite{wi86b} was later modified in Refs.\ \cite{PTY,AMZ1,AMZ2}, depending on whether a certain picture-changing operator in the action was chiral and local \cite{AMZ1,AMZ2} or nonchiral and bilocal \cite{PTY,AKBM,AJK}. A nonpolynomial version of open super-SFT was devised by Berkovits \cite{Ber95,BSZ}. The nonlocal effective action for the tachyon, representing an unstable brane, has been constructed only for the nonchiral version \cite{PTY,AKBM,AJK}. In that case, at lowest order the interaction is slightly more complicated than that of Eq.\ \Eq{rst}:
\be\label{exa}
U(\phi)=\rme^{\B_\Lambda}\phi\,\rme^{\B_\Lambda}\phi^2.
\label{Berkovits}
\ee
Classical solutions of cubic open super-SFT can be mapped into those of Berkovits's nonpolynomial SFT \cite{FK08,Kro92}, so we can regard the nonlocality \Eq{exa} as a generic feature of open superstring field theory at the effective level.

The degrees of freedom corresponding to the graviton are included in the spectrum of closed SFT. Consider the bosonic case \cite{Zwi92,SZ,KKS,KuS,KS3,SZ1,SZ2,SZ3,OZ1,YZ1,YZ2,Mic06,Moe1,Moe2,Moe3} and the action
\be
\hspace{-.4cm}S = \int \! \frac{1}{\alpha^\prime} \Phi \star Q b_0^- \Phi + g \sum_{N = 3}^{+ \infty} 
\frac{ (g \, \alpha^{\prime} )^{N-3}  }{2^{N-3} N !} 
\Phi \star [ \Phi^{N-1}]\,,
\ee
where $\Phi$ is the closed-string field, $Q$ is the BRST operator, $b_0^- = (b_0 - \bar{b}_0)/2$ is a combination of antighost zero modes, and the symbol $[\Phi^{N-1}]=[\Phi_{1}\star\dots\star\Phi_{N-1}]$ represents the string field obtained combining $N-1$ terms $\Phi_{1},\dots,\Phi_{N-1}$ using the $N$-string vertex function. Again, for physical processes dominated by light states, one can truncate the string field in terms of oscillators and particle fields, $\Phi = \Phi_{\rm phys} + \Phi_{\rm aux}$, where
\be
\Phi_{\rm phys} = {c}_0^- \! \left( \phi + A_{\mu \nu} \, \alpha_{-1}^{\mu}  \bar{\alpha}_{-1}^{\nu} \right)  \! | 0 \rangle\,;
\ee
$c_0^- = c_0 - \bar{c}_0$ is a combination of the left and right ghost zero modes; $\phi$ is the bosonic tachyon field; $\alpha_{-1}^{\mu}$ and $\bar\alpha_{-1}^{\mu}$ are the level-($-1$) bosonic Virasoro oscillators; the symmetric and the antisymmetric parts of the tensor $A_{\mu \nu}$ are, respectively, the graviton $h_{\mu \nu}$ and a 2-form $B_{\mu \nu}$; and $\Phi_{\rm aux}$ is made of various auxiliary fields. After gauge fixing and a rearrangement of the auxiliary fields, the resulting effective Lagrangian for the physical modes contains a kinetic and an interaction part, $\mathcal{L} = \mathcal{L}_{\rm free} + \mathcal{L}_{\rm int}$. The kinetic and mass terms read
\be
&& 
\hspace{-0.5cm}
\mathcal{L}_{\rm free} = -\frac{1}{2} \partial_{\lambda} A_{\mu \nu} \partial^{\lambda} A^{\mu \nu} 
- \frac{1}{2} \partial_{\lambda} \phi \partial^{\lambda} \phi - \frac{2}{\alpha^\prime} \phi^2\,,
\ee 
while there are $\sim50$ cubic interaction terms,
\be
&&\hspace{-0.5cm}
 \mathcal{L}_{\rm int} = \frac{g}{3 ! \epsilon^3} \tilde{\phi}^3 +\frac{g  \alpha^{\prime}}{ 2^3 \epsilon^2}  (\partial \tilde{\phi})^2 
\tilde{A}+ \frac{g}{2 \epsilon}  \tilde{\phi} \, \tilde{A}^2 + \frac{g \, \alpha^{\prime} }{2^4 \epsilon} \, \tilde{\phi} \, ( \partial \tilde{A})^2 \nonumber\\
&& \hspace{0.6cm}
 +\frac{g   \, \alpha^{\prime 2} }{2^7 \epsilon} \, \tilde{\phi} \, (\partial^2 \tilde{A})^2 -\frac{ g  \, \alpha^{\prime}}{2^3} \, \tilde{A}^2  
   (\partial^2 \tilde{A}) + \dots \,, \label{Stringint}
\ee
where $\epsilon=2^4/3^3$ and fields with a tilde are ``dressed'' by the same nonlocal operator \Eq{rst} except that the constant $1/\Lambda^2$ has an extra $1/2$ prefactor, which can be removed by a coordinate rescaling.

Supersymmetric versions of closed SFT, both heterotic and Type II \cite{OZ2,BOZ,JM,Mat13}, have been studied less, but one expects to find the same nonlocal structure in effective spacetime actions, Eqs.\ \Eq{rst}, \Eq{exa}, and \Eq{Stringint}. The same is true for a more complete setting including both open and closed strings, as proposals for the $p$-adic tachyonic action indicate \cite{BF2,MoS,Vla06,Ohm03,BCK1}. We conclude that, after a field redefinition \cite{ohm01,KS1,KS2,KS3}, the effective field actions on target spacetime stemming from string field theory have the following schematic structure for bosonic and fermionic fields $\tilde f_i$ (e.g., Refs.\ \cite{BMS,Kos11} and references therein):
\be
S_{\rm eff}= \sum_{i}\int d^D x \left[\frac{1}{2}\tilde f_i (\B-m_i^2)\rme^{-\B_\Lambda} \tilde f_i -U\right],
\label{ESFT}
\ee
where $U$ is a potential (cubic in open SFT) of the fields $\tilde f_i$ in the bosonic case and, in supersymmetric cases such as Eq.\ \Eq{exa}, of the $\tilde f_i$'s further dressed by nonlocal operators. This is the point at which we will connect with quantum-gravity field theories.

From a field-theoretical point of view, the type of nonlocality in Eq.\ \Eq{ESFT} is particularly benign. Not only is the Cauchy problem well defined in general \cite{BK1,cuta3},\footnote{The Cauchy problem can be rigorously defined also for the gravitational theory presented here, as we will show elsewhere \cite{CaMo3}.} but it also allows one to find exact perturbative solutions (i.e., for free fields \cite{BK1}) or very accurate approximate nonperturbative solutions that can go, at the level of target spacetime, beyond marginal deformations \cite{cuta7,roll,cuta5}. These properties are a direct reflection of the gauge symmetries of SFT \cite{cuta7}. All known exact solutions in (open) SFT are superpositions of surface states which obey a ``diffusion'' equation involving only Virasoro and ghost operators. This diffusion equation performs a change of gauge that reparametrizes a trivial non-normalizable solution of the equation of motion into a nontrivial normalized solution. At the level of spacetime fields, the same structure survives in Eq.\ \Eq{ESFT} and, in fact, a spacetime diffusion equation holds for nonperturbative approximate solutions \cite{cuta7}.


\subsection{Quantum field theories of gravity}

In this section, we review a class of gravitational theories constituting a ``nonpolynomial'' extension of the renormalizable quadratic theory by Stelle \cite{Ste77,BOS}. In $D$ dimensions, the Lagrangian has the general structure \cite{Kra87,Tom97,Efi77,MMN,Mod1,BGKM,Mod2,Mod3,Mod4,Mod5}
\be
&& 
 \mathcal{L} =  \frac{2}{ \kappa^{2}} R- R_{\mu \nu}  \gamma_2(\Box_\Lambda)  R^{\mu \nu} \! 
 - R \,\gamma_0(\Box_\Lambda)  R  \nonumber
\\
&& \nonumber \\
&& \hspace{0.75cm}
 - R_{\mu \nu \rho \sigma} \gamma_4(\Box_\Lambda) R^{\mu \nu \rho \sigma}\,,
  \label{theory}
\ee
where $\kappa^2 = 32 \pi G_D$ is the Newton constant, the form factors $\gamma_{0,2,4}$ are entire functions of the ratio $\Box_\Lambda=\B/\Lambda^2$, and $\Lambda$ is an invariant mass scale.

The path to reach Eq.\ \Eq{theory} began with polynomial Stelle's theory in four dimensions, later generalized to $D$-dimensional spacetimes. The Lagrangian, with at most $X$ derivatives of the metric, is
\be
&& \hspace{-0.5cm}
{\mathcal L}_{\rm poly} =  a_2 R + a_4 R^2 + b_4 R_{\mu \nu}^2 + \dots + a_X R^{\frac{X}{2}} + b_X R_{\mu \nu}^{\frac{X}{2}} \nonumber\\
&& \hspace{0.8cm}
+c_X R_{\mu \nu \rho \sigma}^{\frac{X}{2}} + d_X R \, \Box^{\frac{X}{2} - 2} R+ \dots \,,\label{stel}
\ee
where the first dots include a finite number of extra terms with fewer derivatives of the metric tensor and the second ones indicate a finite number of operators $O(R^2 \Box^{X/2-3} R)$ with the same number of derivatives but higher powers of the curvature. Tensor indices are not contracted explicitly and we use the compact notation $R_{\mu \nu}^2:=R_{\mu \nu}R^{\mu \nu}$, and so on. An elementary power counting shows that the maximal superficial degree of divergence of a Feynman graph is $\de= D - (D - X)(V  -  I) = D + (D - X)(L  -  1)$, where $L$, $V$, and $I$ are, respectively,  the number of loops, vertices, and internal lines of a graph. For $X = D$, the theory is renormalizable since the maximal divergence is $\de=D$, and all the infinities can be absorbed in the operators already present in the Lagrangian \Eq{stel}. Unfortunately, however, the propagator contains at least one ghost (i.e., a state of negative norm) that represents a violation of unitarity. 

To overcome this problem, one considers a more general action combining curvature tensors and covariant derivatives of the curvature tensor and containing also nonpolynomial terms \cite{ALS, Salles:2014rua}:
\begin{widetext}
\be
&& \hspace{-1.4cm}
S = \sum_{n=0}^{N+2} \alpha_{2 n} \Lambda^{D - 2 n} \int d^D x \sqrt{|g|}\,\mathcal{O}_{2 n} (\partial_{\rho} g_{\mu \nu}) + S_{\rm NP}  \nonumber  \\
&&  \hspace{-1.05cm}
=  \int d^D x \sqrt{ | g |} \, \big[\alpha_{0} \Lambda^{D} \lambda +  \alpha_{2} \Lambda^{D - 2} R + 
+\Lambda^{D - 4} ( \alpha_{4}^{1}R^2 +  \alpha_{4}^{2}R_{\mu \nu} R^{\mu \nu} +  \alpha_{4}^{3}R_{\mu \nu \rho \sigma} R^{\mu \nu \rho \sigma}) 
\nonumber \\
&& \hspace{-0.75cm}
+ \Lambda^{D - 6} (\alpha_{6}^{1}R^3_{\dots} + \alpha_{6}^{2}\, \nabla R_{\dots} \nabla R_{\dots} + \dots ) 
+ \Lambda^{D - 8} (\alpha_{8}^{1} R^4_{\dots} + \alpha_{8}^{2}  R_{\dots} \nabla R_{\dots} \nabla R_{\dots}+ \alpha_{8}^{3 }\nabla^2 R_{\dots} \nabla^2 R_{\dots} + \dots ) \nonumber  \\
&& \hspace{-0.75cm} 
+ \dots   
+ \Lambda^{D - 2(N+2)} ( \alpha_{2N +2}^{1}R^{N+2}_{\dots}  
+ \alpha_{2N +2}^{2}R^{N-1}_{\dots} \nabla R_{\dots} \nabla R_{\dots}+ \alpha_{2N +2}^{3} R_{\dots} \Box^N R_{\dots}+ \dots ) 
\nonumber \\
&& \hspace{-0.75cm}
+ R_{\mu \nu} \,  h_2( - \Box_{\Lambda}) R^{\mu \nu} + R \, h_0( - \Box_{\Lambda}) \, R\big].  \label{generalAction}
\ee
\end{widetext}
In Eq.\ \Eq{generalAction}, $\mathcal{O}_{2n} (\partial_{\rho} g_{\mu \nu})$ denotes general-covariant operators containing $2n$ derivatives of the metric $g_{\mu \nu}$, while $S_{\rm NP}$ is a nonpolynomial action defined in terms of two entire functions $h_{2,0}$ that we will fix later \cite{Tom97}. The maximal number of derivatives in the local part of the action is $X=2N +4$. From the discussion of the local theory \Eq{stel}, renormalizability sets $2N +4 \geqslant D$. To avoid fractional powers of the d'Alembertian operator, we take $2N+4= D$ in even dimensions and $2N+4 = D+1$ in odd dimensions.

For $N \geqslant 0$ and $n \geqslant 2$, only the operators $R_{\mu \nu} \Box^{n-2} R^{\mu \nu}$, $R \Box^{n-2} R$, and $R_{\mu \nu \alpha \beta } \Box^{n-2} R^{\mu \nu \alpha \beta}$
contribute to the graviton propagator (with $2n$ derivatives in total), but using the Bianchi and Ricci identities the third can be eliminated. To leading order in the curvature, the Lagrangian density in \Eq{generalAction} reads \cite{Mod1,ALS}
\be 
&& \hspace{-0.3cm}
\mathcal{L}  =   -\bar{\lambda}+\frac{2 R}{\kappa^2} - \sum_{n=0}^{N} \left[a_n \, R  (-\Box_{\Lambda})^n R  + b_n  R_{\mu \nu} \, (-\Box_{\Lambda})^n  R^{\mu \nu} \right] \nonumber \\
&&  \hspace{0.4cm}
- R_{\mu \nu} \, h_2( - \Box_{\Lambda}) \, R^{\mu \nu} - R   h_0( - \Box_{\Lambda}) R + O(R^3) \,, \label{action} 
\ee
where we have renamed the constants in Eq.\ \Eq{generalAction} as $\bar\lambda=-\alpha_0 \Lambda^D\lambda$, $2\kappa^{-2}=\alpha_2\Lambda^{D-2}$, and so on. Finally, the entire functions $h_{0,2}(z)$ in Eq.\ \Eq{generalAction} are (here $z:= - \Box/\Lambda^2$) 
\be
&& \hspace{-0.5cm}
h_2(z) = \frac{2[V(z)^{-1} -1]- \kappa^2 \Lambda^2\, z \sum_{n=0}^N \tilde{b}_n \, z^n}{\kappa^2 \Lambda^2\, z} \,,\label{hzD1} \\ 
&& \hspace{-0.5cm}
h_0(z) = -\frac{V(z)^{-1} -1 + \kappa^2 \Lambda^2 \, z \sum_{n=0}^N \tilde{a}_n \, z^n}{\kappa^2 \Lambda^2 \, z} \,,\label{hzD2} 
\ee
for general parameters $\tilde{a}_n$ and $\tilde{b}_n$, where $V(z)$ is a nonpolynomial entire function without zeros in the whole complex plane. 
The couplings and the nonlocal functions of the theory have the following dimensions in mass units: $[\tilde a_n] =[\tilde b_n]= M^{D-4}$, $[\kappa^2] = M^{2-D}$, $[\bar{\lambda}]=M^D$, $[h_2] = [h_0] = M^{D-4}$.
 
The Lagrangian can be expanded at the second order in the graviton fluctuation $g_{\mu \nu} =  \eta_{\mu \nu} + \kappa h_{\mu \nu}$ around the Minkowski background. Adding a gauge-fixing term $\mathcal{L}_{\rm GF}$ due to a local gauge symmetry under infinitesimal coordinate transformations \cite{Ste77}, the linearized gauge-fixed Lagrangian reads 
$\mathcal{L}_{\rm lin} + \mathcal{L}_{\rm GF} = 2^{-1} h^{\mu \nu} \mathcal{O}_{\mu \nu, \rho \sigma} h^{\rho \sigma}$. Inverting the operator $\mathcal{O}$ \cite{AAM}, one finds the following two-point function in the harmonic gauge ($\partial^{\mu} h_{\mu \nu} = 0$) and in momentum space:
\beq\label{propa}
\mathcal{O}^{-1}(k) = -\frac{P^{(2)}}{k^2\bar{h}_2(k^2)} +\frac{P^{(0)}}{(D-2) k^2 \bar{h}_0(k^2)}\,,
\eeq
where
\be
&& \hspace{-0.7cm}
\bar{h}_2(k^2) = 1 + \frac{k^2 \kappa^2 \beta(k^2)}{4}\,,\\
&& \hspace{-0.7cm}
\beta(k^2)     = 2\sum_{n = 0}^{N} b_n (k^2/\Lambda^2)^n + 2\,h_2(k^2/\Lambda^2)\,,\\
&& \hspace{-0.7cm}
\bar{h}_0(k^2) = 1- k^2 \frac{D \kappa^2 \beta(k^2) +4(D-1) \kappa^2 \alpha(k^2)}{4(D-2)},\\
&& \hspace{-0.7cm}
\alpha(k^2)    =   2\sum_{n = 0}^{N} a_n (k^2/\Lambda^2)^n + 2 \, h_0(k^2/\Lambda^2)\,.
\ee
In Eq.\ \Eq{propa}, we omitted the tensorial indices of the operator $\mathcal{O}^{-1}$ and of the projectors $P^{(0)}$ and $P^{(2)}$, defined as \cite{AAM,VN} $P^{(2)}_{\mu \nu \rho \sigma}(k) = (\theta_{\mu \rho} \theta_{\nu \sigma} + \theta_{\mu \sigma} \theta_{\nu \rho})/2 -\theta_{\mu \nu} \theta_{\rho \sigma}/(D-1)$, $P^{(0)} _{\mu\nu\rho\sigma} (k) = \theta_{\mu \nu} \theta_{\rho \sigma}/(D-1)$, $\theta_{\mu \nu} = \eta_{\mu \nu} -k_{\mu} k_{\nu}/k^2$.
 
We now assume that the theory is renormalized at some scale $\mu_0$. Setting
\be
\tilde{a}_n = a_n(\mu_0) \,, \qquad \tilde{b}_n = b_n(\mu_0)\,, \label{betaalphaD}
\ee
the bare propagator only possesses the gauge-invariant, physical massless spin-2 graviton pole, and $\bar{h}_2 = \bar{h}_0 = V^{-1}$. Choosing another renormalization scale, the bare propagator would acquire poles that cancel with a shift in the self-energy in the dressed propagator. Thus, Eq.~\Eq{propa} reads 
\beq 
\mathcal{O}^{-1}(k)= -\frac{V(k^2/\Lambda^2)}{k^2} \left( P^{(2)} - \frac{P^{(0)}}{D-2}\right)\,. \label{propgeneral}
\eeq
As shown by Efimov \cite{Efi77}, 
a nonlocal field theory is unitary and microcausal, provided the following properties are satisfied by $V(z)$:
\begin{enumerate}
\item[(i)] $V(z)$ is an entire analytic function in the complex $z$ plane, and it has a finite order of growth $1/2 \leqslant \rho < + \infty$, i.e., $\exists b>0,c>0$ such that $|V(z)| \leqslant c \, \rme^{b \, |z|^{\rho}}$.
\item[(ii)] When ${\rm Re}(z) \rightarrow + \infty$ ($k^2 \rightarrow + \infty$ or $k_{\rm E}^2 \rightarrow + \infty$), $V(z)$ decreases quite rapidly: $\lim_{ {\rm Re}(z) \rightarrow +\infty} |z|^N |V(z)| =0$, $\forall \, N>0$ [for example: $V(z) = \exp(-z^l)$ for $l \in \mathbb{N}_+$].
\item[(iii)] $[V(z)]^{*}= V(z^*)$ and $V (0) = 1$.
\item[(iv)] The function $V^{-1}(z)$ is real and positive on the real axis, and it has no zeros on the whole complex plane $|z| < + \infty$. This requirement implies that there are no gauge-invariant poles other than the transverse massless physical graviton pole.
\end{enumerate}

\subsubsection{Super-renormalizable quantum gravity}

We now specialize to a form factor $V(z) = \exp (-z^l)$ (where $l\in\mathbb{N}_+$) satisfying the above properties. The high-energy propagator takes the form $\mathcal{O}^{-1}(k) = -\exp[-(k^2/\Lambda^2)^l]/k^2$. The $m$-graviton interaction has the same scaling, since it can be written schematically as ${\mathcal L}^{(m)} \sim h^m \, \Box_{\eta} h \,\exp[(-\Box_{\eta})^l/\Lambda^{2l}]h + \dots$, where $\Box_{\eta} = \eta^{\mu \nu} \partial_{\mu} \partial_{\nu}$ and ``$\dots$'' are other subleading interaction terms. Placing an upper bound to the amplitude with $L$ loops, we find
\be
&& 
``\mathcal{A}^{(L)}   \sim   \int (d^D k)^L \, \left[ \rme^{-(k^{2}/\Lambda^{2})^l} \, k^{-2} \right]^I \, 
\left[\rme^{(k^{2}/\Lambda^{2})^l} k^2 \right]^V \nonumber  \\
  && \hspace{1.0cm}
  =  \int (dk)^{D L} \left[ \rme^{-(k^{2}/\Lambda^{2})^l} \, k^{-2} \right]^{ L-1}"  \, .\label{diverExp}
\ee
In the last step, we used the topological identity $I = V+L-1$. The $L$-loops amplitude is ultraviolet finite for $L >1$, and it diverges at most as ``$(k)^D$'' for $L=1$. Only one-loop divergences survive and, therefore, this theory is super-renormalizable and unitary, as well as microcausal \cite{Kra87,AE1,AE2,Efi77,Efi01}. 
A more rigorous power-counting analysis will be done in a separate paper \cite{LML}. For a form factor $V = \exp [-H(z)]$, where $H(z)$ is an entire function with logarithmic asymptotic behavior, power-counting renormalizability has been shown already in Refs.\ \cite{Tom97, Mod1}. Only a finite number of constants are renormalized in the action (\ref{action}), i.e., $\kappa$, $\bar{\lambda}$, $a_n$, $b_n$, and the finite number of couplings that multiply the operators in the next-to-last line of Eq.\ \Eq{generalAction}. On the other hand, the functions $h_i(z)$ are not renormalized. To see this, we write the entire functions as a series, $h_i(z) = \sum_{r=0}^{+\infty} c_r z^r$. Because of the superficial degrees of divergence $\delta \leqslant D$ derived in Eq.\ \Eq{diverExp}, there are no counterterms that renormalize $c_r$ for $r>N$. Therefore, the nontrivial dependence of the entire functions $h_i(z)$ on their argument is preserved at the quantum level. After multiplicative renormalization, the action becomes
\be 
&& \hspace{-0.1cm} S = \!\int d^D x \sqrt{|g|} \Big\{2\, Z_{\kappa} \, \kappa^{-2} \, R - Z_{\bar{\lambda}} \bar{\lambda} \nonumber \\
&& -\sum_{n=0}^{N} \Big[Z_{a_n} \, a_n \, R \, (-\Box_{\Lambda})^n \, R   + Z_{b_n} \, b_n \, R_{\mu \nu} \, (-\Box_{\Lambda})^n \, R^{\mu \nu} \Big]\nonumber \\
&& -R  \, h_0( - \Box_{\Lambda}) \, R -R_{\mu \nu} \, h_2( - \Box_{\Lambda}) \, R^{\mu \nu} \, \nonumber \\
&& -\sum_{n=1}^{N} Z_{c_n^{(1)}} c_n^{(1)} \, R^{2+n} +  \dots \Big\}. \label{actionRen}
\ee 

\subsubsection{Finite quantum gravity} 

What we have said in the previous section applies to spacetimes of any dimension $D\geqslant 3$, but in this paper we are interested in defining a ultraviolet completion of $11$-dimensional supergravity. In odd dimensions, there are no counterterms for pure gravity at the one-loop level in dimensional regularization, and the theory results in being {\em finite}; it stays so also when matter is added to fill up the supergravity multiplet \cite{duff,DeSe}. We conclude that the amplitudes with an arbitrary number of loops are finite and all the beta functions
vanish,
\beq
\beta_{\kappa} = \beta_{\bar \lambda}=\beta_{a_n} = \beta_{b_n} = \beta_{c_i^{(n)}} = 0\,,
\eeq
where $n =1, \dots, N$ and $i$ runs from 1 to the number of invariants of order $N$. In particular, we can fix all the coefficients $c_i^{(n)}$ to zero, while the couplings $a_n(\mu)$ and $b_n(\mu)$ do not run with the energy: $a_n(\mu) = \tilde{a}_n= {\rm const}$, $b_n(\mu) = \tilde{b}_n = {\rm const}$. Using Eqs.\ \Eq{hzD1} and \Eq{hzD2}, the gravitational Lagrangian density \Eq{action} simplifies to
\be\label{gabox0}
\mathcal{L} = \frac{2}{\kappa^{2}}\left[R+ G_{\mu\nu} \, \gamma(\Box) \,  R^{\mu\nu} \right],
\ee
where we set the cosmological constant to zero and
\be\label{gabox}
\gamma(\Box) :=  \frac{ \rme^{-\Box_{\Lambda}} -1}{\Box} \,.
\ee
The equations of motion up to order $O(R^2)$ are \cite{MMN}
\be
\rme^{-\Box_{\Lambda}} \, G_{\mu \nu} + O(R^2) = 8 \pi G_D \,  T_{\mu \nu} \, . 
\ee


\section{Self-complete geometric unification}\label{5to4d}

In traditional unification scenarios, gauge and gravity fields descend from a higher-dimensional, local, purely gravitational action via a Kaluza--Klein compactification. A negative feature of such theories is that divergences worsen with the increase of the spacetime dimensionality. A different situation arises in nonlocal models. Namely, both renormalization and unitarity of the $D$-dimensional theory are naturally inherited by, say, its $(D-1)$-dimensional compactification. As a simple example of this mechanism, in this section we study the Kaluza--Klein compactification of the action (\ref{action}) on a circle from $D=5$ to $D=4$. One can check that the compactification from $D=11$ to $D=10$ follows similar lines. In this section, we shall indicate with capital Roman letters the indices in $D$ dimensions and with Greek letters those in the reduced spacetime. Let us start with the following purely gravitational action in $D=5$:
\be
\hspace{-0.2cm}
S_{5} = \frac{1}{2 \kappa_{5}^2 }\int d^{5} x \sqrt{ |g |} \left[R + G_{A B } \gamma(\Box)\, R^{A B} \right].  \label{S5} 
\ee
We can expand the five-dimensional line element as
\be  
\hspace{-0.5cm} ds^2 &=& g_{A B} d x^A d x^B\nonumber     \\
&=& g_{\mu \nu} dx^\mu d x^{\nu}  + 2 a A_{\mu} d x^{\mu} d x^5 + a A_{\mu} A_{\nu} d x^{\mu} d x^{\nu} \nonumber     \\
&& + a (d x^5)^2  ,\label{5to4} 
\ee
where $a$ is a dimensionless constant, $A,B = 0,1,2,3,5$ and
$\mu, \nu= 0, \dots,3$. 
The curvature decomposes into
\be
&& \hspace{-0.5cm}
 R^{(5)}_{\mu\nu} = R_{\mu\nu} + O(a A \nabla F) + O(  a F^2) + O(a^2 A^2 F^2) \, , \nonumber\\
&&  \hspace{-0.5cm}
R^{(5)}_{5 5} = \frac{1}{4} a^2 F_{\mu \nu} F^{\mu \nu} \,  ,  \nonumber \\
&&  \hspace{-0.5cm}
R^{(5)}_{\mu 5} = - \frac{1}{2} a \nabla_{\lambda} F^{\lambda}_{ \mu}+\frac{1}{4} a^2 A_{\mu} F_{\nu\rho} F^{\nu\rho}  \,  , \nonumber \\
&&  \hspace{-0.5cm}
R^{(5)} = g^{A B} R^{(5)}_{A B} = R - \frac{1}{4} a F_{\mu \nu} F^{\mu \nu}
\label{deco2}
\,,
\ee
where $F_{\mu\nu}:=\p_\mu A_\nu-\p_\nu A_\mu$.
Let us now expand the action (\ref{S5}) using (\ref{deco2}) and 
omitting all the operators $O(F^4)$, $O(R \nabla A F)$, and
$O(A^4 F^4)$:
\be
\hspace{-.8cm}  S &=&  \frac{1}{2 \kappa_{5}^2 }\int \! d x^5 \!  \int  \! d^{4} x \sqrt{a} \sqrt{|g|} \Big[ R - \frac{1}{4} a F_{\mu \nu} F^{\mu \nu} \nonumber  \\
\hspace{-.8cm}  && + G^{\mu\nu} \gamma(\Box)\, R_{\mu \nu} + 2 R^{(5)\mu 5} \gamma(\Box)\, R^{(5)}_{\mu 5} + {\rm vertices}  \Big]. \label{cinquec}
\ee
Up to extra vertices, the last term reads
\be
\hspace{-0.5cm} 2 R^{(5)\mu 5} \gamma(\Box)\, R^{(5)}_{\mu 5} &=& \frac{2}{a} \left( - \frac{1}{2} a \nabla_{\lambda} F^{\lambda \mu} \right) \nonumber \\
&&\times\gamma(\Box)\left( - \frac{1}{2} a \nabla_{\tau} F^{\tau}_{ \mu} \right) . \label{figo}
\ee
Integrating several times by parts and using the Bianchi identity $\nabla_{\rho} F_{\mu \nu} + \nabla_{\nu} F_{\rho \mu} +\nabla_{\mu} F_{\nu \rho} =0$ for the curvature $F_{\mu\nu}$, the operator (\ref{figo}) turns into
\be
2 R^{(5)\mu 5} \gamma(\Box)\, R^{(5)}_{\mu 5} &=& - \frac{a}{4} F_{\mu \nu} \left( \rme^{ - \Box_{\Lambda} } -1 \right)  F^{\mu\nu}\nonumber \\
&&+ {\rm vertices}\,.\label{FgF}
\ee
Replacing (\ref{FgF}) in (\ref{cinquec}), the local $F^2$ terms cancel each other, and we finally get 
\be
&& \hspace{-0.6cm}
S = \frac{1}{2 \kappa_{5}^2 } \!  \int \! dx^5 \! \int d^{4} x \sqrt{a} \sqrt{|g|} \Big[ R 
+ G^{\mu\nu} \gamma(\Box)\, R_{\mu \nu}  \nonumber\\
&& \hspace{2cm}
- \frac{a}{4} F_{\mu \nu} \rme^{ - \Box_{\Lambda} }  F^{\mu\nu}+ {\rm vertices}
\Big]\,.
\label{quattro}
\ee
The derivative order of the extra vertices is no greater than that of the other operators
in the action. Since the operatorial structure of the form factors is preserved, the theory keeps being super-renormalizable even after compactification. 


\section{M-theory from quantum-gravity oxidation}\label{mth}


We start by considering ten-dimensional Type-IIA supergravity, and then we include the modifications suggested by string field theory while maintaining general covariance. Next, we show how to modify $11$-dimensional supergravity in order to get the previous action by dimensional reduction. To simplify the analysis, we only consider the bosonic sector of the theories. This is legitimate if we assume the existence of a supersymmetric extension of the nonpolynomial $11$-dimensional or $10$-dimensional pure gravity theories, since at the quantum level supersymmetry does not worsen the counterterms of bosonic gravity \cite{Des99}. Actually, in $11$ dimensions it is not so obvious that there are invariant counterterms with an odd number of derivatives. One could imagine an invariant counterterm containing the field strength $F_{\mu \nu \rho \sigma}$ raised at an odd power, but there is a discrete symmetry (time reversal together with $A_{\mu \nu\rho} \rightarrow - A_{\mu \nu\rho}$) that excludes these operators \cite{duff}. 

\subsection{Action}

The main feature of effective actions of string field theory is the exponential damping factor in Eq.\ \Eq{ESFT}. {This form factor belongs to the class of operators independently discussed below Eq.\ \Eq{propgeneral} in the context of nonlocal quantum gravity. We propose to use the knowledge from the graviton action in string field theory to actually fix $\gamma(\B)$ uniquely in the class of nonlinear gravitational actions reviewed in the previous section. The result is Eq.\ \Eq{gabox}, and the action selected out of the infinite class \Eq{theory} is Eq.\ \Eq{gabox0}.}

\begin{figure}
\begin{center}
\includegraphics[height=4cm]{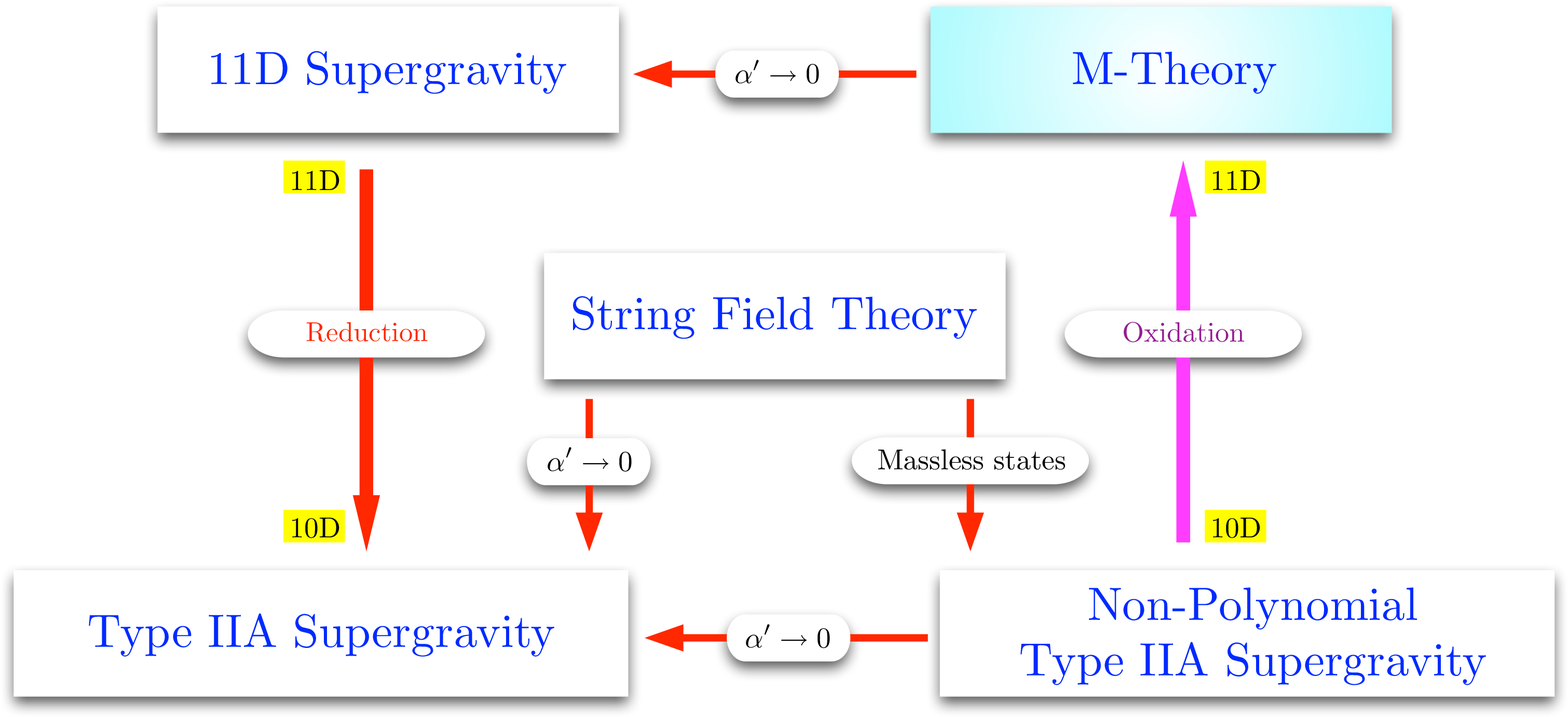}
\end{center}
\caption{\label{fig1} Diagrammatic representation of the completion of $11$-dimensional supergravity. At the core of the graph is string field theory, whose limit $\alpha^{\prime} \rightarrow 0$ is Type-IIA ten-dimensional supergravity. The projection of SFT on the massless sector is nonpolynomial or semipolynomial Type-IIA supergravity. We know that the dimensional reduction of $11$-dimensional supergravity to ten dimensions is Type-IIA supergravity plus extra massive states; applying the reverse process, we can oxidate nonpolynomial Type-IIA supergravity to get a nonpolynomial completion of $11$-dimensional supergravity (``M-theory''). Taking the $\alpha^{\prime} \rightarrow 0$ limit in the M-theory action, we find again the unique $11$-dimensional supergravity. In parallel, the $\alpha^{\prime} \rightarrow 0$ limit of nonpolynomial Type-IIA supergravity is the usual Type-IIA supergravity.}
\end{figure}

Now we introduce the same modification for the other fields that fill up the bosonic sector of supergravity. In $10$ dimensions, we are led to write the action
\be
\hspace{-.2cm} S_{10}^{\rm IIA} &=& \int \frac{d^{10} x}{2 \kappa_{10}^2} \sqrt{|g|}\Bigg\{ \rme^{-2 \Phi} \Big[R+ G_{\mu \nu} 
\frac{\rme^{ - \Box_{\Lambda} } -1}{\Box} R^{\mu \nu}  \nonumber \\
&&  
+ 4 \partial_\mu \Phi \, \rme^{ - \Box_{\Lambda} } \, \partial^{\mu} \Phi- \frac{1}{2 \times 3!} H_{\mu \nu \rho} \, \rme^{- \Box_{\Lambda} } \, H^{\mu \nu \rho} \Big] \nonumber \\
&& 
-\frac{1}{2} \left(\frac{1}{2 !} F_{\mu \nu } \, \rme^{ - \Box_{\Lambda}} \, F^{\mu \nu } + \frac{1}{4 !}  \tilde{F}_{\mu \nu \rho \sigma}  \rme^{ - \Box_{\Lambda} } \tilde{F}^{ \mu \nu \rho \sigma} \right) \Bigg\} \nonumber \\
&& 
- \frac{1}{4 \kappa_{10}^2 } \! \int \!\! B_2 \wedge F_4 \wedge F_4 + \dots \,, 
\label{S10}
\ee
where $\Phi$, $H_3=d B_2$, $F_2$, and $F_4$ are, respectively, the dilaton, the Neveu--Schwarz field strength, and the Ramond--Ramond 2-form and 4-form field strengths, while $\tilde F_4=dA_3-A_1\wedge H_3$ and $\B_\Lambda$ is metric covariant (it does not contain gauge fields). To be in agreement with string field theory, we have to fix $\Lambda^2$ as in Eq.\ \Eq{rst}. We know that 10-dimensional supergravity can be derived from 11-dimensional supergravity by dimensional reduction. Similarly, the theory (\ref{S10}) can be derived from an extension of 11-dimensional supergravity by an oxidation process to be understood as the inverse of Kaluza--Klein dimensional reduction. From what we learned in the previous section, the quantum finite 11-dimensional theory incorporating the SFT propagator as well as general covariance reads
\be
\hspace{-0.4cm}
S_{11} &=& \! \frac{1}{2 \kappa_{11}^2 }  \!  \int \! d^{11} x \sqrt{ |g |} \Big[ R + G_{\mu \nu } \frac{\rme^{ - \Box_{\Lambda} } -1}{\Box} R^{\mu \nu}  \nonumber\\
 \hspace{-0.4cm} && -\frac{1}{2 \times 4 !} F_{\mu \nu \rho \sigma}  \rme^{ - \Box_{\Lambda} } F^{ \mu \nu \rho \sigma} \Big] \nonumber \\
 \hspace{-0.4cm} && - \frac{1}{12 \kappa_{11}^2 } \! \int \!\! A_3 \wedge F_4 \wedge F_4 + \! \dots . \label{S11} 
\ee 
In Eqs.\ \Eq{S10} and \Eq{S11}, we have left the Chern--Simons term untouched, but it should contain also nonlocal operators. Here, we concentrated on the modifications in the gravitational sector and do not discuss the possible modifications to the Chern--Simons action. Figure \ref{fig1} shows the interrelations between local and nonlocal theories under dimensional reduction and oxidation.

It is easy to show that the compactification on a circle of the theory $S_{11}$ gives $S_{10}^{\rm IIA}$, if the Kaluza--Klein states are thrown away. Following Ref.\ \cite{Mod5}, the reader can easily convince themselves that the actions above in 10 and 11 dimensions can be rendered supersymmetric at least up to the quadratic order, because the form factor does not change the spectrum of the theory. A preliminary analysis of supersymmetry has been done for $N=1$ supergravity in four dimensions and $N=1$, $N=2$ supergravity in three dimensions \cite{Mod5}. In that reference, supersymmetry was realized for a nonpolynomial higher-derivative gravity in $N=1$, $D=4$ filling up the supergravity multiplets, paying attention to avoid extra poles in the propagators \cite{dWF}.  

Of course, the extension of Type-IIA supergravity presented in Eq.\ \Eq{S10} does not mimic all the features of string theory but only a general-covariant projection onto the massless sector. This is actually why the theory in ten dimensions is not finite but super-renormalizable. In string theory, the infinite tower of massive particles is responsible for the cancellation of the infinities in the loop amplitudes. However, the compactification of the 11-dimensional finite ``M-theory'' \Eq{S11} should imply a finite ten-dimensional supergravity as well, when the Kaluza--Klein states are included \cite{duff}. 


\subsection{Consistency check of the conjecture}\label{coch}

One way to verify the consistency of our conjecture is to look at the Type-II string theory amplitudes
\cite{W11} (see also Ref.\ \cite{Mag97}). From the four-graviton amplitude in Type-II string theory, there follows the effective action
\be
\tilde S &=& \int d^{10} x \sqrt{|g|} \left( R + R^4 \right) , \label{R4} \\
R^4 &:=& t^{\mu_1 \dots \mu_8}  t^{\nu_1 \dots \nu_8} R_{\mu_1 \mu_2 \nu_1 \nu_2} R_{\mu_3 \mu_4 \nu_3 \nu_4} \nonumber\\
&& 
\times R_{\mu_5 \mu_6 \nu_5 \nu_6} R_{\mu_7 \mu_8 \nu_7 \nu_8},
\nonumber 
\ee
where the symbol $t^{\mu_1 \dots \mu_8}$ is defined so that 
\be
\sqrt{{\rm det} \, \Gamma^{\mu \nu} F_{\mu \nu} }= t^{\mu_1 \dots \mu_8} F_{\mu_1 \mu_2} \dots F_{\mu_7 \mu_8},
\ee
and $\Gamma$ and $F$ are, respectively, the Lorentz generator and the Yang--Mills field strength. Although the Lagrangian (\ref{Stringint}) for closed bosonic SFT includes other interaction vertices with four and six derivatives, these operators are excluded in Type-IIA and Type-IIB string theory, and the first correction to Einstein gravity is $O(R^4)$. However, the action (\ref{R4}) is not unique. Under a local field redefinition $g_{\mu \nu} \to g_{\mu \nu} +  X_{\mu \nu}$,
\be\label{r4}
\tilde{S} = \int d^{10} x \sqrt{|g|} ( R + G_{\mu \nu} X^{\mu \nu} + R^4 + \dots)\,,
\ee
plus higher-order terms. In Ref.\ \cite{W11}, $X_{\mu \nu}$ was taken as a local functional of the metric and its derivatives. Here, we make the particular choice 
\beq \label{notr}
\hspace{-0.1cm} X_{\mu \nu}= \gamma(\Box) R_{\mu \nu}=\left( -1+\frac{\Box}{2 \Lambda^2} - \frac{\Box^2}{6 \Lambda^4} + \dots \right) \! R_{\mu \nu} \,,
\eeq
so that the action (\ref{r4}) agrees with the nonpolynomial action (\ref{S10}), eventually expanded in the form factor. Although we can conclude that our proposal is compatible with the superstring-theory effective action at the first order in the local truncation of the entire function, this result is somewhat weaker than a full physical equivalence. In fact, it is well known that the S-matrix of any Lagrangian field theory is invariant under field redefinitions that are nonlinear but local \cite{Haa58,SaSd,BMSV}. To the best of our knowledge, no analog of this theorem is known to hold for transformations with nonlocal operators such as \Eq{notr}, for which only a comparison with the local theory after truncation of the form factor is presently available.


\subsection{S-duality}\label{sdua}

All closed string theories defined in ten dimensions are invariant under the rescaling 
\be
\Phi &\rightarrow& \Phi - C \, , \nonumber \\ 
g_{s} &\rightarrow& \rme^{C} g_{s} \,,
\ee
together with a possible rescaling of the other fields. Here, $\Phi$ is the dilaton, $g_s$ is the string coupling, and $C$ is an arbitrary constant. Using this rescaling, one can reabsorb $g_s$ in $\Phi$; then, the coupling constant in string theory is related to the dilaton expectation value $\langle \Phi \rangle$. There is no potential for $\Phi$ at the perturbative level, and hence $\langle \Phi \rangle$ can take an arbitrary value. 

The effective theory is also invariant under the rescaling 
\be
\alpha^\prime &\rightarrow& \lambda \, \alpha^\prime \,  , \nonumber \\ 
g_{\mu \nu} &\rightarrow& \lambda \, g_{\mu \nu}\,,
\ee
again together with a rescaling of the other fields. This invariance makes it possible to absorb
 $\alpha^\prime$ into the definition of the metric $g_{\mu \nu}$. As a further check that the nonlocal theory under inspection has viable properties, we now show that it respects S-duality between Type-IIA theory and $11$-dimensional supergravity. When compactifying the latter to ten dimensions, one gets the low-energy limit of nonpolynomial Type-IIA string theory. 
 Upon dimensional reduction, the 11-dimensional metric  $g_{MN}$ 
 decomposes into a scalar $\sigma$, a vector 
 $A_\mu$, and the ten-dimensional metric,
 \be && 
\hspace{-0.2cm} 
ds^2 = g_{MN} d x^M d x^N   \label{deco}  \\
 && \hspace{0.4cm} 
 =g_{\mu\nu}(x^{\mu}) dx^\mu d x^\nu + \rme^{2 \sigma(x^\mu)} [dx^{10}+ A_\nu(x^{\mu}) dx^\nu]^2,
 \nonumber  
 \ee
 where $M,N=0, \dots, 10$ and $\mu, \nu=0,\dots,9$. 
Replacing this decomposition into the 11-dimensional theory, we obtain
 \be
 \hspace{-0.2cm}
  S_{\rm II A} &=& \frac{1}{2 \kappa^2}\int d^{10} x \sqrt{|g|} \rme^\sigma\,
  \left[ R + G_{\mu \nu} \frac{\rme^{- \ell_{11}^2 \Box} - 1 }{\Box} R^{\mu\nu}\right. \nonumber \\
   \hspace{-0.2cm} &&
 - \frac{1}{2 \times 4!} F_{4} \rme^{ - \ell_{11}^2 \Box } F_{4}  - \frac{1}{2 \times 3!} \rme^{- 2 \sigma}H_{3} \rme^{ -\ell_{11}^2 \Box } H_{3}
   \nonumber\\
\hspace{-1.0cm} &&
 \left.- \frac{1}{4} \rme^{2 \sigma}F_{2} \rme^{-\ell_{11}^2 \Box } F_{2} \right]
- \frac{1}{4 \kappa^2}   \int  B_2 \wedge F_4 \wedge F_4 + \dots  ,\nonumber\\\label{tdac}
 \ee
where we introduced the fundamental length $\ell_{11} =1/M_{11}= 1/\Lambda$ in 11 dimensions. 
 In ten dimensions, the gravitational coupling constant for all string theories is given by
 \be
  2 \kappa^2 = (2 \pi)^7 \, \ell_s^8 \, g_s^2\,.
 \ee 
 The gravitational couplings in 11 and 10 dimensions are related by \cite{Kiritsis:2007zza}
 \be
 2 \kappa_{11}^2 = (2 \pi R) 2 \kappa_{10}^2 = (2 \pi)^8 \, \ell_s^9 \, g_s^3,
 \ee
 where $R$ is the radius of the 11th direction 
 and $\ell_s^2 = \alpha^{\prime}$.
 Defining the 11-dimensional Planck mass as $ 2 \kappa_{11}^2 = (2 \pi )^8 M_{11}^{-9}$, we get
 \be
 && \ell_{11} = g_s^{\frac{1}{3}} \ell_s \, , \\
 && g_s = (M_{11} R)^{\frac{3}{2}} \, , \\
 &&  \ell_s = \frac{\ell_{11}^3}{R} \, .
 \label{l11}
 \ee
As in the local case \cite{HuTo,Wit95}, the 11D theory is the nonperturbative decompactification limit $g_s\to\infty$ ($R\to\infty$) of the 10D one. In fact, making use of the metric rescaling
 \be
 g_{\mu \nu} \rightarrow \rme^{ - \sigma} g_{\mu \nu} \quad {\mbox{with}} \quad \sigma = \frac23\Phi \,, 
 \label{stframe}
 \ee
 the action \Eq{tdac} becomes
 \begin{widetext}
     \be
 && S_{\rm II A}  = \frac{1}{2 \kappa^2}  \! \int \!\! d^{10} x \sqrt{|g|} \, \rme^{- 2 \Phi} \!
  \Bigg[ R + G_{\mu \nu} \frac{\rme^{-\alpha^{\prime}  g_s^{\frac{2}{3}}  \rme^{\frac{2}{3} \Phi}  \Box} - 1 }{\Box} R^{\mu\nu} 
 + 4 (\nabla^\mu \Phi)\, \rme^{ -\alpha^{\prime} g_s^{\frac{2}{3}} \rme^{\frac{2}{3} \Phi}\Box }  \nabla_\mu \Phi - \frac{1}{2 \times 4!} F_{4} \rme^{ -\alpha^{\prime}  g_s^{\frac{2}{3}}  \rme^{\frac{2}{3} \Phi} \Box } F_{4} \nonumber \\
&&  \hspace{2.0cm}
 - \frac{1}{2 \times 3!} H_{3} \rme^{ -\alpha^{\prime}  g_s^{\frac{2}{3}}  \rme^{\frac{2}{3} \Phi} \Box } H_{3} 
 - \frac{1}{4} F_{2} \rme^{-\alpha^{\prime} g_s^{\frac{2}{3}} \rme^{\frac{2}{3} \Phi} \Box } F_{2} \Bigg]  
 - \frac{1}{4 \kappa^2} 
  \int  B_2 \wedge F_4 \wedge F_4 + \dots \,,
 \ee
where the dots also include other dilaton contributions coming from the d'Alembertian in the exponential operator between the field strengths.
This is the expression in the Jordan or string frame. Using (\ref{l11}) and the further rescaling $g_{\mu\nu} \rightarrow \rme^{\Phi/2} g_{\mu\nu}$, the 
action in the Einstein frame reads
  \be
&& S_{\rm II A} = \frac{1}{2 \kappa^2}\int d^{10} x \sqrt{|g|} 
  \Bigg[ R + G_{\mu \nu} \frac{\rme^{-\alpha^{\prime}  g_s^{\frac{2}{3}}  \rme^{\frac{\Phi}{6}}  \Box} - 1 }{\Box} R^{\mu\nu}
  + 4 (\nabla^\mu \Phi)\, \rme^{ -\alpha^{\prime} g_s^{\frac{2}{3}} \rme^{\frac{\Phi}{6}}   \Box }  \nabla_\mu \Phi - \frac{1}{2 \times 4!} F_{4} \rme^{ -\alpha^{\prime}  g_s^{\frac{2}{3}}  \rme^{\frac{\Phi}{6}}  \Box } F_{4}
  \nonumber \\
 &&   \hspace{2cm}
 - \frac{1}{2 \times 3!} H_{3} \rme^{ -\alpha^{\prime}  g_s^{\frac{2}{3}}  \rme^{\frac{\Phi}{6}}  \Box } H_{3} 
 - \frac{1}{4} F_{2} \rme^{-\alpha^{\prime} g_s^{\frac{2}{3}} \rme^{\frac{\Phi}{6}}  \Box } F_{2} \Bigg]  
 - \frac{1}{4 \kappa^2} 
  \int  B_2 \wedge F_4 \wedge F_4 + \dots \, .
  \label{IIAE}
 \ee
 \end{widetext}
As an application of the scale invariance common to all string theories,  
we explicitly give the transformation rules for fields and couplings of Type-IIA supergravity. 
The action (\ref{IIAE}) is invariant under the transformations
\be
g_{\mu \nu} &\rightarrow& \rme^{C/2} g_{\mu\nu} \,, \nonumber\\
\Phi &\rightarrow& \Phi - C \, , \nonumber \\
g_{s} &\rightarrow& \rme^{C} g_{s}\,, \nonumber \\
F_{2} &\rightarrow&  \rme^{C} F_{2} \,  , \nonumber \\
F_{4} &\rightarrow&  \rme^{C} F_{4} \,  .
\ee
This invariance makes it possible to fix $\alpha^{\prime} = g_s =1$.

 


\section{Effective action including massive states}\label{massi}

In this section, we consider the infinite tower of massive states present in the 
string spectrum and we formally introduce the full effective action compatible with
string field theory and general covariance.
Apart from the massless sector, the ten-dimensional theory (\ref{S10}) should include all the massive states. 
At quadratic order, the Lagrangian density reads
\be
\hspace{-.4cm} \mathcal{L}_{\rm massive} &=& 
 \sum_{n=1}^{+\infty} 
\phi^{(n)}_{\mu_1 \dots \mu_k} (\Box - M_{n}^{2}) \, \rme^{- \Box_{\Lambda}} \, \phi^{(n) \mu_1 \dots \mu_k}
 \\
 && 
+ \mathcal{V} \left(\phi^{(1)}, \dots, \phi^{(\infty)}, g_{\mu \nu},B_{[2]}, \Phi, H_{[3]}, A_{[1]} \right) \!, \nonumber
\ee
where the potential $\mathcal{V}$ includes all the interactions with the massless states
as well as between the massive states compatible with string field theory. 

We first truncate the infinite tower to a fixed integer number $N^*$ and calculate the one-loop effective action $\Gamma^{(1)}$. Here, we are interested in the beta functions coming from the divergent contribution to $\Gamma^{ (1) }$. Integrating out all the massive states, the effective action is
\be
\hspace{-0.1cm} \Gamma^{(1)} &=& - \rmi  \ln Z  \label{Gamma1}\\
  &=& S_{\rm II A}(\bar{g}) + S_{\rm massive}(\bar{g}) + \frac{\rmi}{2}   \ln \det (\hat{H}_{\rm II A})\nonumber \\
  && +  \frac{\rmi}{2} \sum_{n=1}^{N^*}   \ln \det (\hat{H}_{\phi^{(n)}})- \rmi  \ln\det (\hat{M})- \frac{\rmi}{2} \ln\det (\hat{C})  \, ,\nonumber 
\ee
where we employed the background field method and decomposed the metric as $g_{\mu \nu} = \bar{g}_{\mu \nu} + h_{\mu\nu}$. 
The determinants for the operators $\hat M$ and $\hat C$ come from the ghost contributions, 
$\hat{H}_{\rm IIA}$ is the second-order variation of the ten-dimensional action with respect to all the massless fields (including the gauge-fixing term), and 
$\hat{H}_{ \phi^{(i)} }$ is the second-order variation of $S_{\rm massive}$ with respect to the massive fields. In compact notation,
\be
 \hat{H}_{\rm II A } &=& \frac{\delta^2 S_{\rm II A}}{ \delta h_{\mu \nu}\delta h_{\rho \sigma}}\Bigg|_{h=0} 
 + \frac{\delta \chi_{\delta}}{\delta h_{\mu \nu}} \, C^{\delta \tau } \, 
 \frac{\delta \chi_{\tau}}{\delta h_{\rho \sigma}}\Bigg|_{h=0} , \nonumber \\
  \hat{H}_{ \phi^{(n)} } &=& 
 \frac{\delta^2 S_{\rm massive}}{ \delta \phi^{(n)}  \delta \phi^{(n)} }\Bigg|_{ \phi^{(n)}=0}  .
\ee
In defining the effective action for the massless fields, all the external massive states are fixed to zero, and 
the one-loop effective action reads
\be
&& \Gamma^{(1)} =  S_{\rm II A}(\bar{g}) 
+ \frac{\rmi}{2}  {\rm Tr} \ln (\hat{H}_{\rm II A})+ \frac{i}{2} \sum_{n=1}^{N^*}{\rm Tr} \ln \left[  \hat{H}_{ \phi^{(n)} } 
  \right]  \nonumber \\
  && \hspace{1cm}
  - \rmi\, {\rm Tr} \ln (\hat{M})
 - \frac{\rmi}{2}{\rm Tr} \ln (\hat{C})  \, .
\ee
Evaluating the traces, we find the divergent contribution
\be
\hspace{-0.2cm} 
\Gamma^{(1)}_{\rm div} &\propto& - \frac{1}{\epsilon} \int \rmd^{10} x \sqrt{|g|}\times   \nonumber \\
&& \hspace{-0.1cm} 
\!\left[  \beta_{R} R +
\beta_{R^2} R^2 + \beta_{{\rm Ric}^2} R_{\mu \nu} R^{\mu\nu} 
+ \dots +\sum_i \beta_{R^5}^{(i)} (R^5)_i \right] \nonumber \\
&& \hspace{-0.1cm} 
+ \sum_j \beta^{(j)} {\mathcal O}_{(j)}(\Phi, B_{[2]}, F_{[2]}, F_{[4]}, {\rm fermions})\,,
\ee
where the index $i$ indicates all possible combinations of curvature invariants of a given order. The other operators ${\mathcal O}_{(j)}(\Phi, B_{[2]}, F_{[2]}, F_{[4]})$ of dimension up to $M^D$ involve the other massless fields, for example $F_{[2]}^2$, $F_{[2]}^3$, $F_{[2]}^4$, $F_{[2]}^5$, $F_{[2]}^2 (\nabla F_{[2]} )^2$, and so on.

The running coupling constants multiply all the operators of dimension up to $M^D$ that we can add to the bare action (\ref{S10}) from the beginning. However, the quantum divergences depend only on the vertices coming from the form factors (the other vertices contribute to the finite part of the effective action). The beta functions $\beta^{(i)}$ turn out to be independent of the running coupling constants but dependent on the nonrunning parameters defining the form factor. 
Contributions to $\beta^{(i)}$ come from the massless and massive states circulating in the one-loop amplitudes, namely, 
\be
&& \beta^{(i)}_{*} := \beta^{(i)}_0 + \sum_{n=1}^{N^*} \beta^{(i)}_n(M_n)=: \ \sum_{n=0}^{N_*} \beta^{(i)}_n(M_n)\,, 
\nonumber \\
&& 
M_0 \equiv 0 \,,
\label{beta10}
\ee
where the massless-states contribution from $\hat{H}_{\rm IIA}$ has been denoted as $\beta^{(i)}_0$. The finite contributions to the Lagrangian density in the ultraviolet regime read 
\be
 \hspace{-0.1cm}
\mathcal{L}_{\rm fin} &\propto& \beta_{R^2} R \ln \left( - \frac{\Box}{\mu^2} \right)  R + \beta_{{\rm Ric}^2} R_{\mu \nu} \ln \left( - \frac{\Box}{\mu^2} \right) R^{\mu\nu} \nonumber \\
&& + \dots +\sum_i \beta_{R^5}^{(i)} \left[R^3 \ln \left( - \frac{\Box}{\mu^2} \right)  R^2 \right]_i \nonumber
\\
&&  \!+\! \sum_j \beta^{(j)} {\mathcal O}_{(j)} \left[ \Phi, B_{[2]}, F_{[2]}, F_{[4]}, {\rm fermions}, \ln \left( -\frac{\Box}{\mu^2} \right) \right]\! ,  \nonumber \\
\ee
where $\mu$ is the renormalization scale. 
 
At this point, there are two possibilities. One is to assume that the beta functions vanish order by order in the string spectrum,
\be
\beta^{(i)}_n(M_n) = 0 \qquad \forall\, n,i \,.\label{MN}
\ee
One can thus achieve finiteness by introducing the minimal number of local operators needed to satisfy (\ref{MN}) at any mass-spectrum level. Such a conjecture could be checked by considering a level-by-level truncation of the action. This is not a trivial task because the exponentially growing form factor is compatible with the presence of many other interaction operators in the action, as long as they do not invalidate super-renormalizability [in other words, these operators in momentum space cannot grow faster than any of the other operators present in (\ref{S10})]. 

However, this order-by-order prescription is not a feature to be expected in string theory, where only the presence of an infinite number of fields makes the amplitudes finite. As we will comment in Sec.\ \ref{fiit}, string theory is possibly finite (and not only super-renormalizable) as a consequence of modular invariance; therefore, we expect the beta functions $\beta^{(i)}$ in (\ref{beta10}) to be identically zero in the limit $N^* \rightarrow + \infty$:
\be
\beta^{(i)} := \sum_{n=1}^{+\infty} \beta^{(i)}_n(M_n)  =0 \quad \forall\, i\,.
\ee
This relation will need to be checked in the future. The massive states in string theory play a similar role as of Pauli--Villars fields in quantum field theory, but with the crucial double difference of being infinitely many and physical (i.e., they do not include ghosts).\footnote{Another way to make a ten-dimensional theory finite is to introduce a (non)local gravitational potential at least cubic in the Riemann tensor. Then, the theory is finite by a suitable choice of the coefficients in the operators characterizing the potential \cite{LML}. At present, this mechanism seems incompatible with the previous one involving an infinite number of massive states because the above-mentioned potential cannot be generated by integrating out the massive states.}



\section{Finiteness in quantum gravity and string theory}\label{fiit}

Evidence is known that, in string theory, loop amplitudes are finite. This has been verified up to two loops and four external states in string scattering amplitudes. Assuming that  such finiteness held for all amplitudes, one would like to understand the reasons beyond this property. Could it be due to the infinite tower of higher-spin fields or to supersymmetry or both? If we focus on the ultraviolet divergences, we see that bosonic string theory itself is finite, although the presence of a tachyon in the spectrum undermines the stability of the theory. The introduction of supersymmetry allows one to remove the tachyon, but it is not able to tame the divergences, as evident in local supergravity. These data highlight the crucial role carried out by higher-spin fields in the string spectrum in making the theory finite. However, in general, it is not sufficient to introduce an infinite number of particles: they also have to fit in a particular mass spectrum. The finiteness of the theory is achieved if the mass spectrum implies a particular symmetry in the loop amplitudes that, in the string case, is modular invariance \cite{Pol98}. This symmetry forces us to exclude the ultraviolet region from the integration domain in the loop amplitudes so that physical processes result in being finite.

Looking at the effective SFT action \Eq{ESFT}, there is an asymmetry between the kinetic and the interaction term, due to the presence of the exponential factor in the quadratic operators. As mentioned in Sec.\ \ref{sft}, the exponential operator is a consequence of the reparametrization symmetries of the full theory. Such asymmetry would imply a suppression of interaction terms in the ultraviolet and suggests that string theory does not manifest ultraviolet divergences, contrary to a general-covariant effective action. However, due to the fact that the Lagrangian is written in terms of curvature invariants, general covariance implies the same nonlocal operators in the kinetic term \emph{as well as} in the interaction vertices for gravity. Therefore, we are apparently able to get a finite theory only at second and higher orders in the loop expansion, but not at the one-loop level, since we find the same scaling both in the graviton propagator and the vertices. On the other hand, general coordinate invariance in string theory can only be achieved through cancellations among contributions from infinitely many interaction terms \cite{KS3}. Therefore, the role of the infinite number of massive states is twofold: to assure modular invariance (via the particular mass spectrum of the string) and diffeomorphism invariance. 

The nonuniqueness of string theory in ten dimensions reflects the compatibility of different mass spectra with both requirements. Let us consider again the gravity sector of the nonpolynomial theory \Eq{S10}. Expanding the action in powers of the graviton field and making the field redefinition
\beq
h := e^{ \frac{\Box_{\Lambda}}{2} } \, \tilde{h} \,, 
\eeq
up to the three-graviton vertex we get the action 
\be
 \hspace{-0.1cm} S &\sim&  \! \int  \! d^D x  \left(h \, \Box   \, \rme^{- \Box_{\Lambda}} \, h+  \, h \, 
\partial^2 h \,  \rme^{- \Box_{\Lambda}}  h  + \dots \right)  \\ 
&=& \!  \int \! d^D x \left[ \tilde{h} \, \Box \, \tilde{h} + \, \left( \rme^{\frac{\Box_{\Lambda}}{2}}  \tilde{h}  \right) \, \partial^2 \left( \rme^{\frac{\Box_{\Lambda}}{2}}  \tilde{h}  \right) 
  \rme^{- \Box_{\Lambda}}  \left( \rme^{\frac{\Box_{\Lambda}}{2} }  h  \right) \right]\!,
  \nonumber 
\ee
which is very close to the structure in (\ref{ESFT}) and even more so to the structure in (\ref{Berkovits}). However, in the present form, the theory is no longer explicitly general covariant. These arguments indicate that, possibly, also effective string field theory is not finite when general covariance is explicitly recovered or implemented. If this was the case, then the perturbative finiteness of string theory (in ten dimensions) would mainly follow from modular invariance. 

We mention a further complication in the picture that illustrates the interplay of various perturbative and nonperturbative ingredients. The finiteness of nonpolynomial supergravity in 11 dimensions, Eq.\ \Eq{S11}, is not a reflection of modular invariance in ten dimensions, since we are not considering higher-spin fields in either theory. Actually, following the considerations in Ref.\ \cite{duff}, also the ten-dimensional theory \Eq{S10} is expected to become finite if we introduced the infinite tower of Kaluza--Klein modes coming from 11D nonpolynomial supergravity compactified on a circle. These states are nonperturbative from the string-theory point of view \cite{Wit95}. 

However, ten-dimensional nonpolynomial supergravity could be made finite also in a different way, by introducing all the higher-spin fields of the perturbative string spectrum. (Other possible states to be introduced could come from the quantum membrane in 11 dimensions \cite{duff}.) Given that in both cases the ten-dimensional theory might become finite, it is permissible to ask whether the two different completions are actually related by a nonperturbative duality. The role of the exponential form factor in this and other dualities should be checked in the future. Another possibility is to make the 10D theory finite by applying the Pauli--Villars regularization scheme \cite{Diaz:1989nx, Anselmi:1991wb, Anselmi:1992hv, Anselmi:1993cu, Modesto:2014xta}. It is indeed well known that the presence of the infinite tower of resonances in string theory acts, in some sense, like a Pauli--Villars regulator. Infrared divergences survive, but they can be controlled by using supersymmetry.
A last option we wish to mention is to introduce an $O(R^3)$ local potential to make all the beta functions vanish \cite{LML}.


\section{Discussion}\label{concl}

In this paper, we began by reviewing an extended theory of gravity capable of overcoming the renormalizability problem of Einstein gravity in the ultraviolet regime. At the quantum level, the theory is super-renormalizable in even dimensions. In odd dimensions, at one loop, there are no local invariant counterterms with an odd number of derivatives, and the theory turns out to be finite. All the other loop amplitudes are finite because of the nonpolynomial nature of the action. Then, we fixed the nonlocal kinetic terms of the theory according to what was found in effective string-field-theory actions and proposed the resulting supersymmetric extension as an $11$-dimensional quantum supergravity, i.e., as a candidate field-theory limit of the massless sector of M-theory. The proposal is consistent with the compactification from 11 to 10 dimensions and has the correct local limit expected both in 11D and Type-II supergravity.

There are several open points that can be explored to verify the feasibility and consequences of the picture. 

$\bullet$ \emph{Quantum theory and supergravity.} Although our proposal satisfies all the properties (a)--(d) simultaneously, an important consistency check would be to carry out the more ambitious program of the Wilsonian effective action mentioned in the Introduction. Until then, our model can have some beneficial role in the advancement of the field by constituting a practical alternative in which several cross-checks and avenues of exploration are readily available. As outlined above, the role of higher-spin states in the UV behavior of the theory is still to be fully assessed. Also, the theory in $11$ dimensions has direct implications for four-dimensional quantum gravity. In $D=4$, there is optimistic evidence that local $N=8$ supergravity be perturbatively finite (see Ref.\ \cite{Kal3} and references therein), but this view is not universally shared \cite{Ban12} and the theory may even have a divergence at seven loops \cite{GRV,BjG,Bjo10}. Compactifying 11D nonpolynomial supergravity to four dimensions would give rise to a nonlocal version of $N=8$ supergravity. Possibly, also the four-dimensional theory would be finite, provided we retain the infinite tower of massive states coming from the Kaluza--Klein compactification. Moreover, we showed that our proposal respects S-duality between Type-IIA supergravity and M-theory. Various other checks of S-duality as in Ref.\ \cite{Sen:1998kr} and of T-duality \cite{GPR,Sch96,BHO} could be devised at this stage.

$\bullet$ \emph{Singularities.} Point (v) in Sec.\ \ref{intro} deserves further attention, too. Explicit solutions, found via the diffusion equation method \cite{cuta7} or perturbative techniques \cite{BK1}, can indicate whether infinities in the theory are removed also in the sense of spacetime singularities. The cosmology of our model in $D=4$ has been considered in Ref.\ \cite{BMMS} in the local limit and in Ref.\ \cite{SFC} exploiting the asymptotic freedom of the theory. In particular, the results of Ref.\ \cite{SFC} indicate that the cosmic big-bang singularity might be indeed removed, and that other nontrivial phenomenology (such as primordial acceleration) is induced by the same nonlocal mechanism, as generally found in nonlocal cosmologies with exponential operators \cite{BMS,cuta8,BKM1,BKMV,BM2}. It is likely that also black holes in M-theory nonlocal gravity would show, perhaps not surprisingly, interesting physics.

{$\bullet$ \emph{$p$-adic string and complex poles.} For completeness, we briefly comment on the arising of $p$-adic features in the quantum theory. $p$-adic models are based on propagators of the form appearing in $p$-adic string theory, which shares some properties with string field theory \cite{W2,BF2,MoS,Vla06,Ohm03,BCK1}. For instance (and setting all couplings to 1), a free $p$-adic scalar field is characterized by a Lagrangian of the form $\cL=\phi(\rme^{-\B}-1)\phi$, i.e., without an overall (polynomial of the) $\B$ multiplying the entire function. Contrary to string field theory, however, such models have a tachyonic mode (not to be confused with the tachyon of the bosonic string). In fact, the dispersion relation associated with the $p$-adic scalar is
\be\label{padicpro}
\rme^{k^2}-1=0\,, 
\ee
whose solutions are $k^2=-(k^0)^2+|{\bf k}|^2=2\pi\rmi n$, where $n\in\mathbb{Z}$. If the spatial momentum is real, then $k^0$ is complex valued, and there is an instability in Lorentzian signature. The latter represents false-vacuum excitations of the $p$-adic string. On the other hand, the form factor typical of string field theory gives rise, for a massless field, to the dispersion relation
\be\label{sftpro}
k^2 \rme^{k^2} =0\,,
\ee
which is not invariant under $k^2 \to k^2+2\pi\rmi n$. Therefore, in the SFT case, there are no extra complex or real poles in the propagator. For this and other reasons ($p$-adic actions are not derived as effective actions from proper string theory), $p$-adic systems are regarded as simplified nonlocal toy models of some more fundamental SFT structure. Nevertheless, they can be employed to obtain some useful insight in SFT.

The theory \Eq{gabox0} with form factor \Eq{gabox} is of SFT type rather than $p$-adic. As we have seen, at quadratic order in the graviton fluctuations the $\B$ in the denominator compensates two of the four derivatives in the numerator, and the outcome is the bare propagator $G(k^2)\sim -\exp (-k^2)/k^2$ around flat spacetime. 
The advancement in the techniques for understanding the dynamics of nonlocal theories does not actually require to use $p$-adic toy models, since such techniques can be applied directly.\footnote{We will do so for the theory presented in this paper in a forthcoming publication \cite{CaMo3}.} However, our theory might possess some $p$-adic features at the quantum level. To be more precise, $p$-adic-like dispersion relations may arise as a result of quantum corrections to the bare propagator. In fact, the theory \Eq{gabox0} is tree-level unitary and, at the quantum level, we do not expect extra real poles apart from those of the local spectrum. However, other operators will be generated in the quantum action and the dressed dispersion relation for the graviton will contain extra $k$-dependent terms with respect to Eq.\ \Eq{sftpro}. In general, the solutions of such a dispersion relation will be complex, similarly to what happens for Eq.\ \Eq{padicpro}. This proliferation of complex conjugate poles has been recently considered in Ref.\ \cite{Sha15}. In our case, however, the theory is finite at the quantum level, and one can show that such complex poles can be removed by a field redefinition with a trivial Jacobian. The actual presence and physical interpretation of these poles deserve further attention.}

$\bullet$ \emph{Extended objects.} Finally, it is worth noticing that nonlocality has been often advocated as a description of {extended objects}. This interpretation is suggested, \emph{inter alia}, by studies on the nonlocal particle action \cite{Kat90,CHL}
\beq\label{ppa}
S_{\rm particle} = \int d\tau \, \dot{x}_{\mu} \, \gamma(\partial_{\tau}) \, \dot{x}^{\mu}\,, 
\eeq
where $\gamma(\partial_{\tau})$ is a form factor in one dimension. In Ref.\ \cite{Kat90}, it was shown that the open string emerges naturally from Eq.\ \Eq{ppa} for a particular choice of the form factor not dissimilar from ours, while in Ref.\ \cite{CM1} we gave a precise determination of the dependence of the dimensionality of the object described by \Eq{ppa} from the resolution scale at which it is observed. In the context of quantum gravitational theories, an entire function with asymptotically polynomial behavior in a region around the real axis is \cite{Tom97}
\be
&&
 V^{-1}(z) =\rme^{H(z)}=\exp\left[{\sum_{n =1}^{+ \infty} \, \frac{(-1 )^{n+1} \, p(z)^{2 n}}{2n  \, n!}}\right]\nonumber \\
&& \hspace{1.23cm}
= \rme^{\frac{1}{2} \left\{\Gamma[0, p(z)^2]+\gamma_{\rm E}+\ln[p(z)^2]\right\}},\nonumber
\ee
where $\gamma_{\rm E}$ is Euler's constant and $p$ is a polynomial of a certain degree. When expanded for small $z$ (small momenta), the function $H$ does not include the linear term of string field theory. To obtain it, here we modify the above expression as $H\to H'$, where
\be
\hspace{-0.0cm}
H'(z) &=& H(z) + 1 - \rme^{-z}  \label{chain} \\
\hspace{-0.4cm}
&=& 1 - \rme^{-z} + \frac{1}{2} \left\{\Gamma [0, p(z)^2]+\gamma_{\rm E}  + \ln [p(z)^2] \right\} \nonumber\\
\hspace{-0.4cm}
&=& \!\!\! 
\underbrace{ 0 }_{\mbox{particle}} \!\!\! + \underbrace{z}_{\mbox{string}} + \dots + \underbrace{z^{l}}_{?} + \dots  + \underbrace{\frac{1}{2} \ln [ p(z)^2]}_{?} .\nonumber 
\ee
With this choice of form factor, the theory is well defined and perturbatively calculable \cite{LML}. Equation \Eq{chain} shows a local point-particle phase, a string phase, and other phases that, in 11 dimensions, one might conjecture to represent extended objects such as membranes as well as high-energy corrections to the particle phase. The spectral dimension $\ds$ can be a useful tool to understand the nature of the underlying objects. For $H'$, $\ds=11$ when $z\to 0$ ($\Lambda \rightarrow + \infty$, IR), $\ds = 0$ in all the higher-derivative intermediate regimes, and $\ds>0$ in the high-energy regime $z\to +\infty$. A possible interpretation of this result is that the theory interpolates between two point-particle regimes feeling different dimensions, passing through multiple phases dominated by extended objects with vanishing extension along one or more directions in the ambient spacetime. However, the relation between nonlocality and extended objects is more delicate than this direct identification. In fact, nonlocal operators do not change the poles of standard local propagators, which means that the string and membrane spectrum of local and nonlocal versions of the theory is the same. In particular, the insertion of nonlocal form factors would not change the spectrum of matrix or noncommutative models in an essential way. The presence of membranes in the 11-dimensional theory is unrelated to nonlocality: it is rather signaled by the presence of $p$-forms in effective target actions (local or nonlocal). However, nonlocal operators could thicken the branes of the local theory, as the results of Ref.\ \cite{CM1} suggest. This effect could be verified, for instance, with the techniques of Ref.\ \cite{CM1} when applied to a brane effective action. Moreover, to check whether M2- and M5-branes actually are included in our theory, one should also verify that metrics representing M2- and M5-branes are solutions of the nonlocal dynamical equations in 11 dimensions (see, e.g., Sec.\ 4.9 of  Ref.\ \cite{133} for the local case). We expect that the brane metric will be singularity free as in the black hole case \cite{MMN}: the exponential form factor would smear the Dirac delta source in the transversal directions into a Gaussian, and the solution would be regular everywhere. To check the existence and validity of such solutions, and whether they are exact or approximate, would require the full equations of motion, which will be presented and studied in Ref.\ \cite{CaMo3}. Preliminary calculations indeed show that M2- and M5-branes are exact dynamical solutions.

These hints are suggestive of the possibility, to be checked in the future, that the class of theories presented in this paper can provide a general framework for a description of extended objects such as strings and membranes, thus strengthening our \emph{tenet} that the action \Eq{1} correctly captures the main degrees of freedom of M-theory in the field-theory limit. 

\section*{Acknowledgments} 

We thank R.\ Brandenberger, H.\ Verlinde, and E.\ Witten for useful discussions and acknowledge the i-Link cooperation programme of CSIC (Project No.\ i-Link0484) for partial sponsorship. The work of G.C.\ is under a Ram\'on y Cajal contract.


\end{document}